\documentclass[aps,twocolumn]{revtex4-2}
\usepackage{graphicx}
\usepackage{amsmath}
\begin{document}

\title{Sensing decoherence by using edge state} 
\author{Andrey R. Kolovsky}
\affiliation{Kirensky Institute of Physics, 660036, Krasnoyarsk, Russia} 
\affiliation{Siberian Federal University, 660041 Krasnoyarsk, Russia} 
\affiliation{Kimyo International University in Tashkent, 100121 Tashkent, Uzbekistan} 
\date{\today}

\begin{abstract} 
In the absence of decoherence the current of fermionic particles  across a finite lattice connecting two reservoirs (leads) with different chemical potentials is known to be ballistic. It is also known that decoherence typically suppresses this ballistic current. However, if decoherence is weak, the change in the current may be undetectable.  In this work we show that the effect of a weak decoherence can be amplified by orders of magnitude  if the lattice has edge states.  
\end{abstract}

\maketitle
{\em 1.}
Since the discovery of the topological insulators the edge states in topologically nontrivial latices have attracted lot of attention in many different systems \cite{book1}. In particular, we mention fascinating laboratory experiments with photonic crystals, where transport along edges of two-dimensional photonic lattice was visualized {\em in situ} \cite{Hafe13,Khan17}.  Besides three- and two-dimensional topological lattices there are also quasi one-dimensional and truly one-dimensional lattices \cite{Su79} which have exponentially localized states located at the lattice ends. As the result of this localization, edge states in one-dimensional lattices do not contribute to quantum transport  and, in this sense, there are no principal difference between lattices with and without edge states. However, it appeared to be not the case if we have decoherence/dephasing processes in the system, for example, due to residual interaction with the system enviroment. The fact that decoherence can essentially modify or even enhance the coherent quantum transport is well known  and the environment-assisted quantum transport was analyzed in many particular systems \cite{Rebe09,Prio10,Bigg16,Skal25}. The advantage of the considered in this work one-dimensional lattices with edge states is their potential application  as a decoherence sensor. We will show below that weak decoherence creates the new conduction window in the gap between Bloch bands where the edge states are located. It is argued that by measuring the stationary current in this window one can measure  decoherence rate in the system with high precision.  

In the paper we analyze the problem where two reservoirs of Fermi  particles with slightly different chemical potentials are connected by the Su-Schrieffer-Heeger (SSH) lattice, Sec.~II, and by the flux rhombic lattice, Sec.~III. We mention, in passing,  that both lattices attract considerable interest and  have been realized in laboratory experiments by using different physical platforms.  In particular, the topological invariant of the SSH lattice -- the Zak phase -- was directly measured by using cold atoms in the double-periodic optical lattice in Ref.~\cite{Atal13} and phenomenon of the Aharonov-Bohm caging in the flux rhombic lattice was directly observed  in the photonic crystal  in Ref.~\cite{Cace22} and in the array of transmons  in Ref.~\cite{bb}.  In the present work we use these two lattices as typical representatives of one-dimensional lattices with edge states. While the SSH lattice is simpler for the numerical analysis the flux rhombic lattice has the unique feature that for the Pieirls phase $\phi$ equal to $\pi$ all eigenstates of this lattice including the edge states are the compact states and, thus, all Bloch bands are flat \cite{Leyk18}. This allows us to obtain in this particular case the analytic estimate for the stationary current in the conduction windows. 

\begin{figure}
\includegraphics[width=8.5cm,clip]{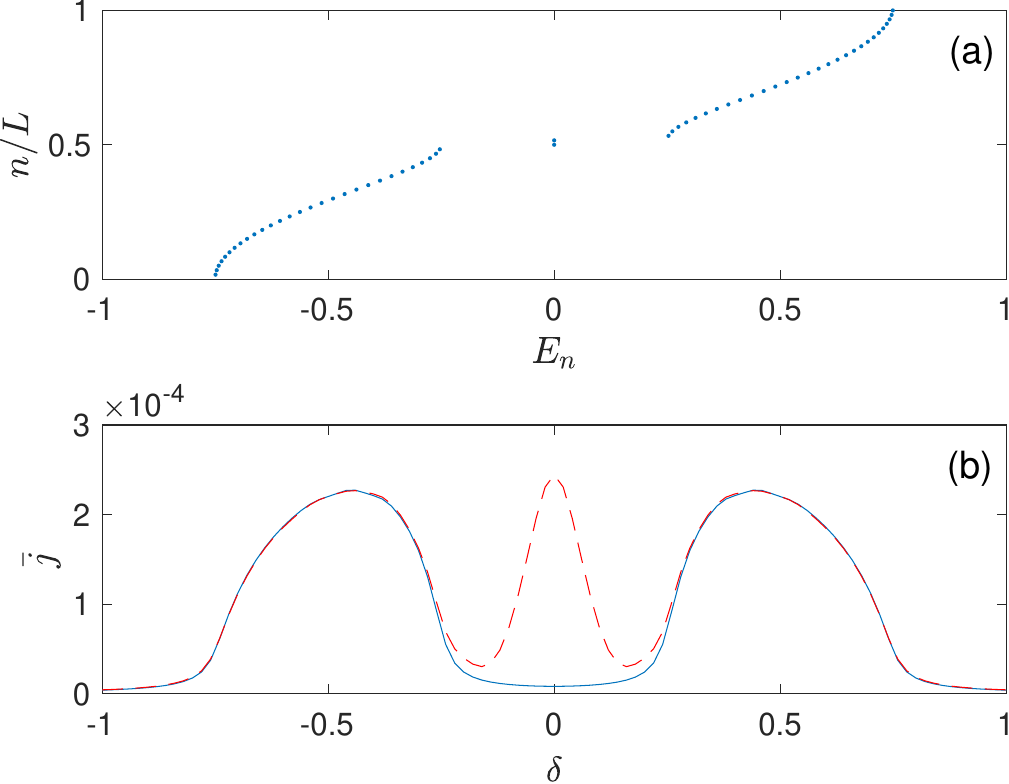}
\caption{Upper panel: Energy spectrum of the SSH lattice of the length $L=60$ with two edge states. The hopping elements are $J=1$ and $\tilde{J}=0.5$. Lower panel: The current across the lattice as the function of the gate voltage $\delta$ for $\kappa=0$ (blue solid line) and $\kappa=0.003$. Parameters of reservoirs are $\mu_{\rm L}=\pi/40$, $\mu_{\rm R}=-\pi/40$, $1/\beta=0$, and $\gamma=0.05$. The coupling constant is $\epsilon=0.2$.}
\label{fig1}
\end{figure}

{\em 2.}
As the first example we consider the SSH lattice,
\begin{equation}
\label{ssh}
 \widehat{H}_{\rm s} = \delta \sum_{\ell=1}^L |\ell\rangle\langle\ell | 
 - \sum_{\ell=1}^{L-1} \frac{J_\ell}{2} \left(|\ell+1\rangle\langle\ell |  +h.c.\right)  \;,
 \end{equation}
where the hopping matrix elements $J_\ell$ take values $J$ and $\tilde{J}\ne J$ for alternating sites. As for any bipartite lattice, the energy spectrum of the SSH lattice consists of two Bloch bands separated by the energy gap. However, due to topological nature of the SSH lattice, the finite SSH of the length $L$ may have edge states with energies located in the gap, see upper panel in Fig.~\ref{fig1}.  The number of edge states depends on how the lattice is terminated and, thus, vary between zero and two.  In what follows we will focus on the case with two edge states.

Now we attach two reservoirs of Fermi particles (the leads) to the lattice ends,
\begin{equation}
\label{ham_tot}
 \widehat{H}_{\rm tot} = \widehat{H}_{\rm s} +\sum_{\rm i=L,R} \left( \widehat{H}_{\rm i} +\epsilon \widehat{H}_{\rm s,i}  \right) \;,
 \end{equation}
where we introduce the coupling constant $\epsilon$. Usually, one models  the lead Hamiltonians $\widehat{H}_{\rm i}$ by semi-infinite tight-binding chains. Alternatively, one can model  them by the tight-binding rings of the size $M$ where $M$ eventually tends to infinity \cite{120,131}.  We are interested in the current across the lattice for $\epsilon\ne0$. The common method to calculate this current is the Landauer theory which relates the system conductance to the transmission probability $|t|^2$ for the plane waves propagating in the leads. The latter is known to exhibit a number of transmission peaks of the width $\Gamma_n\sim \epsilon^2$ located at eigenvalues $E_n$ of the Hamiltonian (\ref{ssh}). Exclusions are  eigenvalues $E_n=0$ associated with the edge states because, instead of resonant transmission, here we meet the phenomenon of resonant trapping characterized by the Wigner delay or dwell time \cite{cc}.  Thus, according to the Landauer theory the stationary current as the function of the parameter $\delta$ should show $L-2$ transmission peaks. In practice, however, these individual peaks may not be resolved because of the peak broadening  due to finite chemical potential difference of the leads and finite temperature. Another source of the transmission peak broadening is the dissipative dynamics of the carriers in the leads \cite{131,Ajis12,Zelo14,Grus16}. 

All above mentioned broadening mechanisms are fully captured by the master equation for the total density matrix ${\cal R}={\cal R}(t)$,
\begin{equation}
\label{master1}
\frac{d {\cal R}}{d t}=-i[\widehat{H}_{\rm tot},{\cal R}] + \sum_{\rm i=L,R} {\cal L}_{\rm i}({\cal R})   \;,
\end{equation}
where the Lindblad relaxation operators ${\cal L}_{\rm i}({\cal R})$ enforce  relaxation of the isolated ($\epsilon=0$) leads to the thermal equilibrium characterized by the inverse temperature $\beta$ and the chemical potential $\mu$ of the respective lead.  In terms of the reduced density matrix of a given, let us say the left reservoir, $\rho_{\rm L}(t)={\rm Tr}_{\rm s,R}[{\cal R}(t)]$, the explicit form of this operator in the quasimomentum basis $| k\rangle$ is  
\begin{equation}
\label{lind}
{\cal L}_{\rm i}[\rho_{\rm i}(t)] = -\gamma[\rho_{\rm i}(t) -\tilde{\rho}_{\rm i}] \;, \quad 
\tilde{\rho}_{\rm i}=\sum_k  n_k |k\rangle\langle k |
\end{equation}
where $n_k$ are the occupation numbers of the quasimomentum states given by the Fermi-Dirac distribution,
\begin{equation}
\label{thermal}
 n_k=\frac{1}{e^{\beta(E_k-\mu)} + 1}      \;,\quad  E_k=-J\cos\left(\frac{2\pi k}{M}\right) \;.
\end{equation}
Notice that operators  ${\cal L}_{\rm i}({\cal R})$ act only on the leads and, thus, the current across the lattice is ballistic independent of the value of the relaxation constant $\gamma$. 

Since our aim is the effect of decoherence on the stationary current across the lattice, we include in the master equation (\ref{master1}) additional relaxation operator
\begin{equation}
\label{decoherence1}
{\cal D}(\rho_{\rm s})= -\frac{\kappa}{2} \sum_{\ell} 
( \hat{\ell}\hat{\ell}\rho_{\rm s}-2\hat{\ell}\rho_{\rm s}\hat{\ell}+\hat{\ell}\hat{\ell}\rho_{\rm s} )  \;,\quad  \hat{\ell}=| \ell \rangle\langle \ell |  \;,
\end{equation}
which acts on the lattice.  To some extent, the operator (\ref{decoherence1}) describes the effect of inter-particle interactions on the single-particle density matrix of the carriers \cite{61}, not mentioning the effect of environment.  It is easy to show that for $\epsilon=0$ this relaxation operator causes exponential decay of off-diagonal elements of the density matrix $\rho_{\rm s}(t)={\rm Tr}_{\rm L,R} [{\cal R}(t)]$ with the rate $\kappa$.  The operator (\ref{decoherence1}) is known to change the ballistic transport regime into the diffusive regime, so that in the thermodynamic limit $L\rightarrow\infty$ the current decreases as $1/L$ \cite{Znid10b}. In what follows, however, we will consider finite lattices where classification of different transport regimes is not so obvious.
    
\begin{figure}
\includegraphics[width=8.5cm,clip]{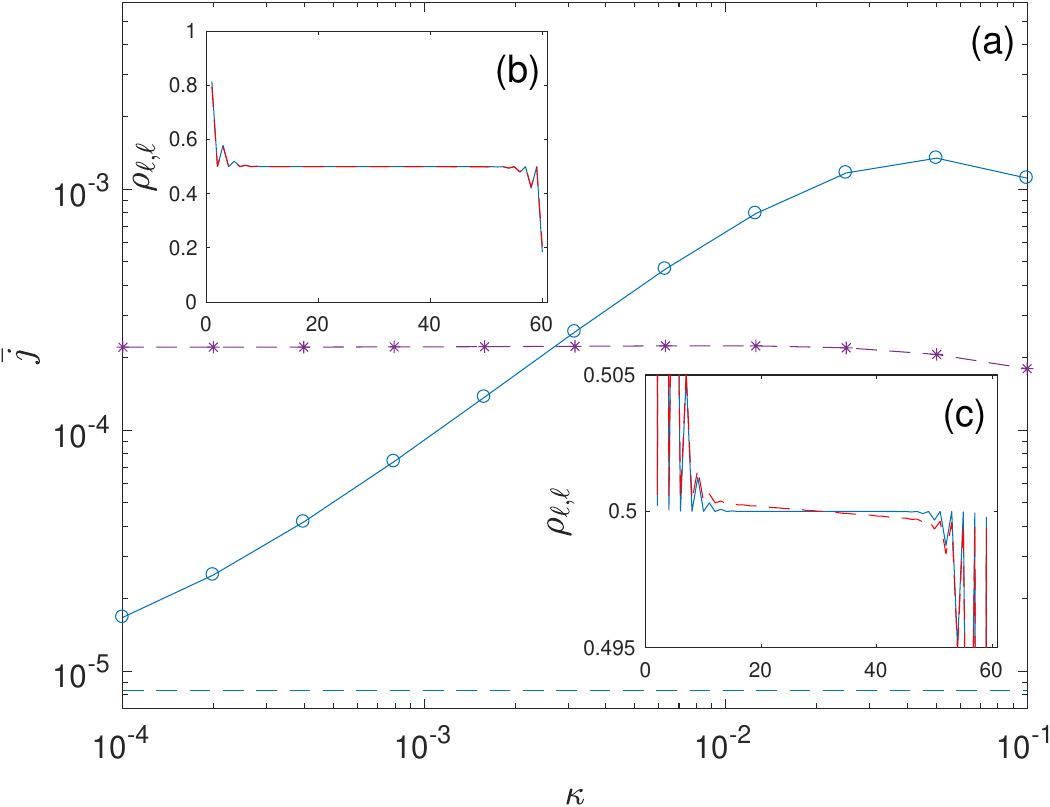}
\caption{Main panel: Stationary current as the function of the decoherence rate $\kappa$ for $\delta=\pm0.4$ (asterisks) and $\delta=0$ (open circles). The dashed lines indicates the current magnitude for $\delta=0$ and $\kappa=0$. Insets: Occupations of the lattice sites in the stationary regime for $\delta=0$ and $\kappa=0$, blue solid line, and  $\kappa=0.003$, dashed red line. The panel (c) zooms in the central region of the panel (b).} 
\label{fig2}
\end{figure}

We proceed with numerical results.  First, we discuss the case $\kappa=0$. The solid line in Fig.~\ref{fig1}(b) exemplifies the stationary current $\bar{j}$ across the SSH lattice in the case where the individual resonant peaks are not resolved. The obtained dependance of the current on the gate voltage $\delta$ obviously reproduces the band structure of the SSH lattice. Notice that the edge states do not participate in the ballistic transport and, thus, cannot be revialed in the dependence $\bar{j}=\bar{j}(\delta)$. Yet, they manifest themselves in the stationary population of the lattice sites. It is seen in Fig.~\ref{fig2}(b) that the presence of edge states results in huge population imbalance between the left and right end sites of the lattice,  which is the consequence of the above mentioned resonant trapping. We stress that this imbalance appears only if the energy interval 
\begin{equation}
\label{interval}
\mu_{\rm R}-\gamma/2<E<\mu_{\rm L}+\gamma/2
\end{equation}
includes  energies of the edge states. Going ahead, we also mention that this huge imbalance is the underlying mechanism for `amplification'  of the decoherence effect in the system.

Let us now $\kappa\ne0$ and let us discuss the destructive effect of decoherence in some more detail. As it was already mentioned, without taking the thermodynamic limit $L\rightarrow\infty$ the identification of different transport regimes is not a trivial task. One of signatures of the diffusive and sub-diffusive currents is nonzero population gradient ${\rm d}\rho_{\ell,\ell}/{\rm d}\ell$ in the central part of a finite lattice \cite{Bert21}. Clearly, this population gradient is a continuous  function of the decoherence rate and it may vary from $+0$ to the maximally possible value $1/L$. Thus, it looks reasonable to distinguish between weakly diffusive and strongly diffusive transport regimes depending on the value of the population gradient. In what follows we focus on the limit of small decoherence rate where  the destructive effect of decoherence is hard to detect and, thus, the current is almost ballistic, see Fig.~\ref{fig2}(c). However, if the lattice has edge states this negligible destructive effect of decoherence is converted into the constructive effect  -- the new conduction window in the energy gap which is easy to detect.

The dashed line in Fig.~\ref{fig1}(b) shows the dependence of the stationary current $\bar{j}$ on the gate voltage $\delta$ for $\kappa=0.003$. The new transmission peak at $\delta=0$ is clearly seen. Since the origin of this peak is the resonant population of the edge states, its width is determined by Eq.~(\ref{interval}) which is independent of $\kappa$. On the contrary, the hight of the peak is mainly determined by the decoherence rate. The dependence of the peak hight on $\kappa$  is depicted by open circles in Fig.~\ref{fig2}(a). Additionally, asterisks show the current in the conductance bands at $\delta=\pm0.4$ and the dashed line indicate the value of the current at $\delta=0$ and $\kappa=0$. (In fact, for $\kappa=0$ this residual current in the band gap can be made arbitrary small by decreasing $\gamma$.)   It is seen in Fig.~\ref{fig2}(a) that for small $\kappa$, where the  destructive effective of decoherence is negligible (see asterisks),  the resonant peak at $\delta=0$ grows linearly with $\kappa$. Thus, by measuring the hight of this peak one can measure the decoherence rate in the system.

\begin{figure}[t]
\includegraphics[width=8.0cm,clip]{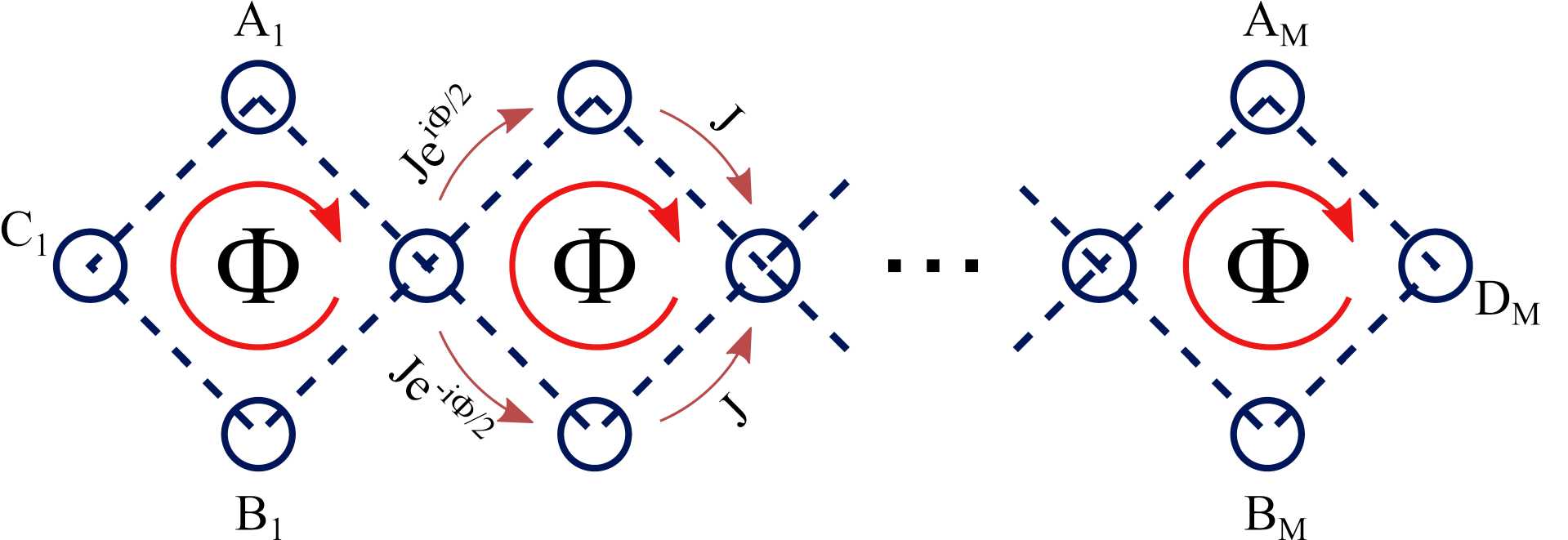}
\caption{The flux rhombic lattice.}
\label{fig0}
\end{figure}

\begin{figure}[b]
\includegraphics[width=8.5cm,clip]{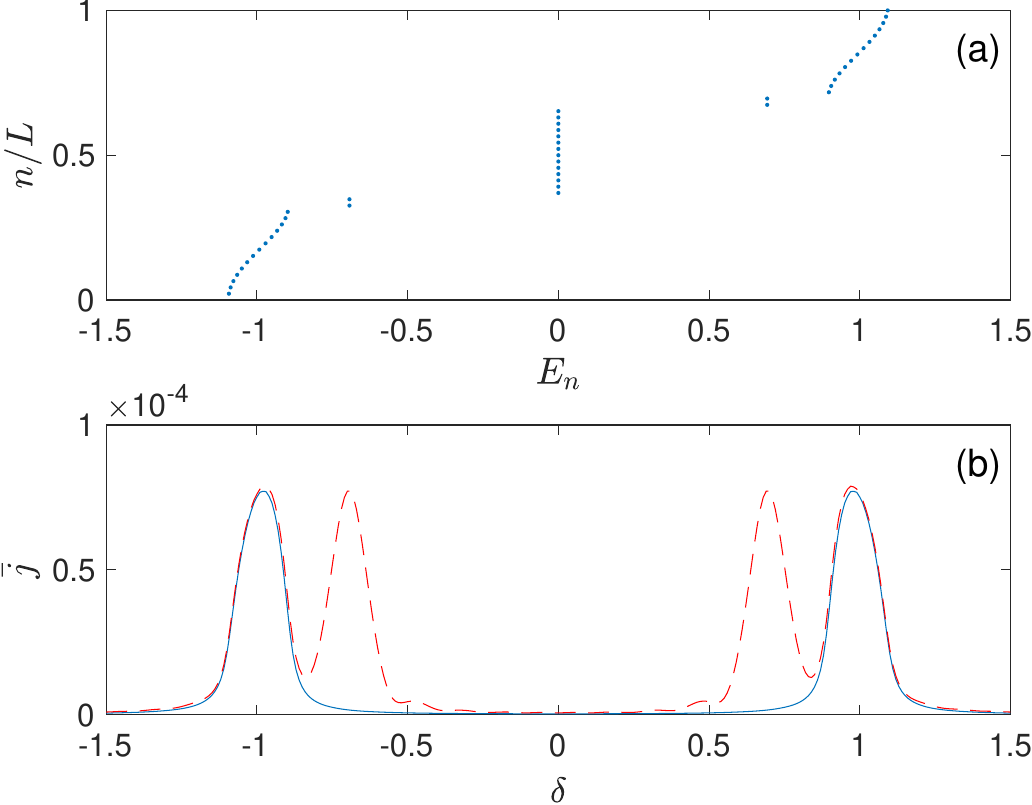}
\caption{Upper panel: Energy spectrum of the flux rhombic lattice with $L=15$ rhombs for $\phi=\pi-0.4$. The modulus of the hopping elements is $|J|=1$ . Lower panel: The current across the lattice as the function of the gate voltage $\delta$ for $\kappa=0$ (blue solid line) and $\kappa=0.001$ (red dashed line). Parameters of reservoirs are $\mu_{\rm L}=\pi/40$, $\mu_{\rm R}=-\pi/40$, $1/\beta=0$, and $\gamma=0.05$. The coupling constant is $\epsilon=0.2$.}
\label{fig1b}
\end{figure}

\begin{figure}
\includegraphics[width=8.5cm,clip]{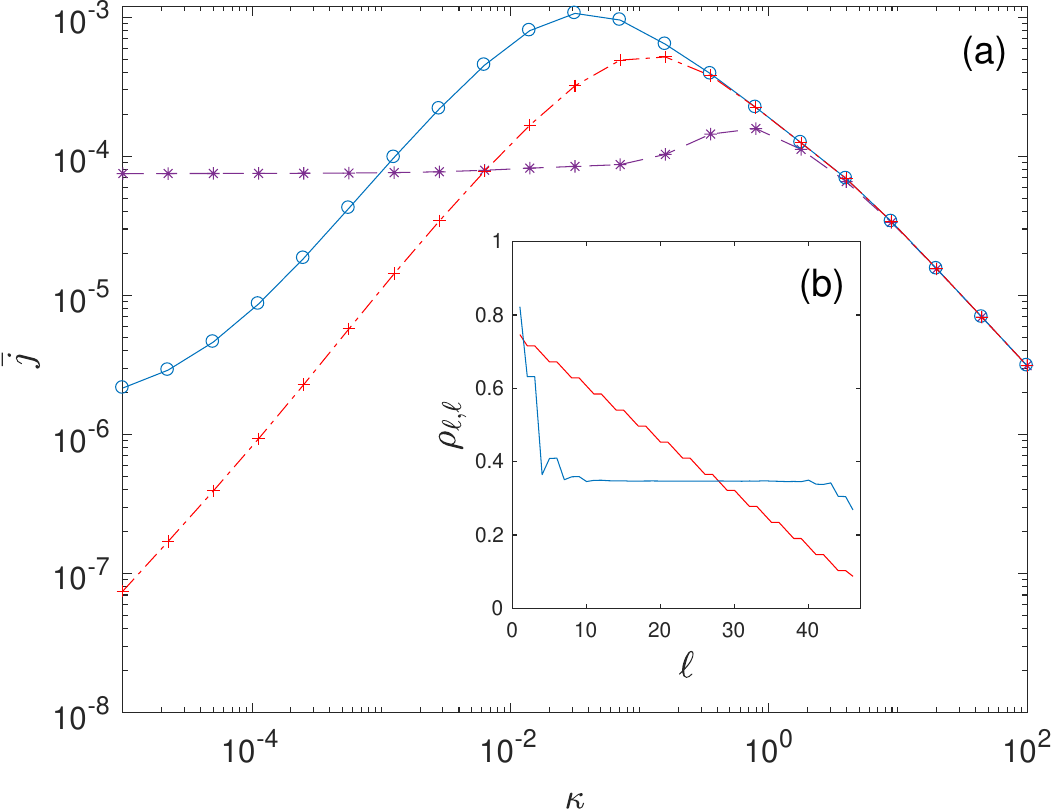}
\caption{Stationary current as the function of the decoherence rate $\kappa$ for $\delta=1$ (asterisks) and $\delta\approx0.7$ corresponding to energies of edge states (open circles). The other system parameters are the same as in Fig.~\ref{fig1b}.  The additional red dash-dotted line is the stationary current for $\phi=0$ where it is pure diffusive. The inset chows the lattice site populations for $\kappa=0.001$ and $\phi=\pi-0.4$ (blue line) and $\phi=\pi$ (red line).}
\label{fig2b}
\end{figure}

{\em 3.}
As the second example we consider the flux rhombic lattice, see Fig.~\ref{fig0}. The control parameter of this lattice is the Peierls phase $\phi$, which in the case of charged fermions is determined by the value of the magnetic flux through a rhomb. If $\phi\ne\pi$ the spectrum of the flux rhombic lattice consists of one flat band, two dispersive bands, and four exponentially localized edge states, see Fig.~\ref{fig1b}(a). Then, it is not surprising that transport properties of the rhombic lattice resemble those of the SSH lattice. In particular, for nonzero decoherence rate $\kappa$ the stationary current across the lattice as the function of the gate voltage $\delta$ shows additional  transmission peaks  at energies of the edge states, see red dashed line in Fig.~\ref{fig1b}(b). The heights of the peaks associated with Bloch bands  and edge states  are shown in Fig.~\ref{fig2b} by asterisks and open circles, respectively.  As compared to Fig.~\ref{fig2}, in Fig.~\ref{fig2b} we extend for the purpose of future discussion the upper limit of the $\kappa$ axis up to $\kappa=100$. Considering for the moment only the region of small $\kappa<0.1$, the results shown  in Fig.~\ref{fig2b} are seen to be similar to those obtained for the SSH lattice. Let us also mention that, according to the population gradient criterion the  transport regime for $\kappa<0.1$ and the considered value of the Peierls phase $\phi=\pi-0.4$ should be classified as weakly diffusive, see blue solid line in Fig.~\ref{fig2b}(a).

As it was already mentioned in Sec.~I, the unique feature of the rhombic lattice is that for $\phi=\pi$ the dispersive bands becomes flat. Thus, the ballistic transport across the lattice is forbidden and the dependence $\bar{j}=\bar{j}(\delta)$ has only two peaks which are associated  with the edge states. Notice that  for $\phi=\pi$ these states are the compact states, where only the sites $C_1$, $A_1$,  $B_1$  or the sites $A_L$, $B_L$, $D_L$ have nonzero occupation numbers.  Remarkably, this qualitative change from the exponentially localized to the compact edge states is reflected in the change from the weakly diffusive transport regime to the strongly diffusive transport regime, see red solid line in Fig.~\ref{fig2b}(b). 

The characteristic feature of the pure diffusive regimes is the Esaki-Tsu like dependence of the stationary current on the decoherence rate $\kappa$,  
\begin{equation}
\label{tsu}
\bar{j}\sim\frac{\kappa}{\kappa^2+const} \;.
\end{equation}
The numerically obtained dependence $\bar{j}=\bar{j}(\kappa)$ in the case $\phi=\pi$ is depicted in Fig.~\ref{fig2b} by the dash-dotted line and it is indeed  can be well approximated by the Esaki-Tsu formula (\ref{tsu}). Let us also mention with this respect that Eq.~(\ref{tsu}) is also in agreement with the analytic result  \cite{max} where the stationary current across the flux rhombic lattice was analyzed within the framework of the Markovian master equation for the boundary driven flux rhombic lattice.

{\em 4.}
We studied the effect of weak decoherence on the quantum transport of fermionic particles in the tight-binding lattices with edge states, namely,  the SSH lattice and the flux rhombic lattice.  For vanishing decoherence rate the edge states do not contribute to the current because of their finite localization length. However these states are resonantly populated/depleted when the Fermi energy of the leads  coincides with the energy of a given edge state. This creates a huge imbalance  in the stationary population of the left and right ends of the lattice, which is a precondition  for amplification of the current induced by decoherence. 

The simplest from the theoretical viewpoint case is the flux rhombic lattice with the Peierls phase $\phi$ equal to $\pi$, where all Bloch bands are flat and the edge states are compact. Here the current induced by decoherence is pure diffusive and, as the function of the decoherence rate, obeys the celebrated Esaki-Tsu equation. The general case, which is actually simpler from the experimental viewpoint, is provided by the SSH lattice,  where Bloch bands are dispersive and the edge states are only exponentially localized. Here a weak decoherence opens the new conduction window in the energy gap between dispersive bands. Interestingly, the current in this window    
is neither pure diffusive nor ballistic. Moreover, its magnitude can exceed the magnitude of the pure ballistic current in the absence  of decoherence processes.  We argued that by measuring this current one can measure very low decoherence rates which would be undetectable if the lattice had no edge states.

\vspace{0.5cm}
The author acknowledges hospitality of the Kimyo International University in Tashkent. 


\end{document}